\documentclass[prc,showpacs,groupedaddress,floatfix,twocolumn]{revtex4}%

\usepackage{graphicx}
\usepackage{amssymb}
\begin{document}

%\begin{frontmatter}

\title{The effective force NL3 revisited}

\author{G. A. Lalazissis$^{1,3}$, S. Karatzikos$^{1,3}$, R. Fossion$^{1}$, D. Pena
Arteaga$^{3}$, A. V. Afanasjev$^{2}$, P. Ring$^{1,3}$}

\address{$^{1}$ Department of Theoretical Physics, Aristotle
University of Thessaloniki, GR-54124, Greece}

\address{$^{2}$ Department of Physics and
Astronomy, Mississippi State University, Mississippi State,
Mississippi 39762, USA}

\address{$^3$ Physik-Department, Technische Universit\"{a}t M\"{u}nchen
D-85747, Garching, Germany}

\begin{abstract}
Covariant density functional theory based on the relativistic mean
field (RMF) Lagrangian with the parameter set NL3 has been used in
the last ten years with great success. Now we propose a modification
of this parameter set, which improves the description of the ground
state properties of many nuclei and simultaneously provides an
excellent description of excited states with collective character in
spherical as well as in deformed nuclei.
\end{abstract}

%\begin{keyword}
%covariant density functional theory\sep relativistic mean field
%theory
%\PACS 21.10.-k\sep 21.60.-n\sep 24.10.Cn\sep 21.30.Fe\sep
%21.60.Jz\sep 24.30.Gz


%\end{keyword}

%\end{frontmatter}
\maketitle

Density functional theory is a universal and powerful tool for
describing properties of finite nuclei all over the periodic table.
In the non-relativistic framework the most successful density
functionals are the ones based on density dependent forces, such as
the Skyrme~\cite{VB.72} or the Gogny~\cite{DG.80} functional.
Relativistic mean field (RMF) theory was first introduced as a fully
fledged quantum field theory by Walecka~\cite{Wal.74,SW.86}. However,
it turned out very soon~\cite{BB.77}, that for a quantitative
description of nuclear surface properties an additional density
dependence is necessary. Nowadays RMF theory modified in this form is
considered as a covariant form of density functional theory. Over the
years it has gained considerable interest, in particular, for the
description of nuclei at and far from stability
\cite{Rin.96,VALR.05,MTZ.06,Pie.07,Fur.07}. Compared with
non-relativistic density functionals covariant density functional
theory has certain advantages. They are characterized by a new
saturation mechanism obtained by a delicate balance between a
strongly attractive scalar field and a strongly repulsive vector
field. Moreover, the very large spin-orbit splitting, observed in
finite nuclei, is a relativistic effect. Therefore, its treatment in
relativistic models arises in a natural way without any additional
adjustable parameters. In addition, time-odd mean fields which are
important in systems with broken time reversal symmetry are uniquely
defined in RMF theory because of the Lorentz covariance of the
underlying Lagrangian~\cite{AR.00}.

Pairing properties are essential for a description of nuclei with
open shells. They have been included first in the constant gap
approximation by occupation numbers of BCS-type~\cite{NL1}. Since
this procedure requires the knowledge of the experimental pairing
gaps, it cannot be applied in unexplored regions of the nuclear
chart, where the binding energies are not known. In addition, it is
noted that the BCS approximation breaks down in nuclei far from the
valley of stability, where the coupling to the continuum is
essential~\cite{DFT.84}. Therefore the constant gap approximation has
been replaced by relativistic Hartree-Bogoliubov (RHB)
theory~\cite{GEL.96} which includes a finite range particle-particle
interaction of Gogny form. The details of this theory have been
discussed in several review articles~\cite{Rin.96,VALR.05,MTZ.06} and
in the references given there.

In any case the adopted functionals are considered universal in the
sense that they can be used for nuclei all over the periodic table,
where mean field theory is applicable. It is therefore very desirable
to find a unique parameterization for the Lagrangian of the model,
which is able to describe as many experimental data as possible. In
other words, we search for an effective force that is able to
describe properties of nuclei from light to very heavy, from the
proton to the neutron drip line. Moreover, a powerful density
functional should not only describe the ground state properties of
finite nuclei but also, at the same time, collective excited states
within time-dependent density functional theory. This is crucial
because otherwise, one would end up with one effective force for the
description of the masses, another one for exotic systems, a third
one for giant resonances an so on. Definitely, such specific
parameterizations do not serve the purposes of universal density
functional theory. Their inability originates from the fact that they
are tailored to describe either specific observables or narrow
regions of the periodic chart.

%%%%%%%%%%%%%%%%%%%%%%%%%%%%%%%%%%%%%%%%%%%%%%%%%%%%%%%%%%%%%%%
\begin{table*}[t]
\centering%
\tabcolsep=3pt%
\renewcommand{\arraystretch}{1.5}%
\caption{{\protect\small The experimental values of the energies of
the Giant Monopole (GMR) and Giant Dipole (GDR) Resonances in
$^{208}$Pb are compared with the predictions of various non-linear
forces.}}%
\vspace{0.5cm}%
\label{Tab1}%
\begin{tabular}
[c]{rrrrrrrr}\hline\hline%
& \hspace{0.7in}Exp. & \hspace{0.4in}NL3 &
\hspace{0.3in}NL3$^{*}$ & \hspace{0.2in}NL-SH & \hspace{0.2in} NL-1 &
\hspace{0.2in}NL-Z & \hspace{0.2in}NL-RA1\\\hline GMR &
$14.1\pm{0.3}$ & $14.04$ & $13.90$ & $16.48$ & $12.44$ & $11.72$ &
$15.26$\\
GDR & $13.3\pm{0.1}$ & $12.95$ & $12.95$ & $12.98$ & $12.97$ &
$12.53$ & $12.87$\\%
\hline\hline
\end{tabular}
\end{table*}
%%%%%%%%%%%%%%%%%%%%%%%%%%%%%%%%%%%%%%%%%%%%%%%%%%%%%%%%%%%%%%%

There are several types of relativistic density functionals.
Conventional RMF theory is based on the Walecka model, where the
nucleons interact by the exchange of phenomenological mesons and, up
to now, in all the successful relativistic models two additional
assumptions are essential: (i) in \textit{the mean field
approximation} only the nucleonic fields are quantized and the
nucleons move independently in classical mesons fields, depending in
a self-consistent way on the nuclear densities and currents, and (ii)
in the \textit{no-sea approximation} vacuum polarization and the
contributions of the negative energy solutions are not explicitly
taken into account.

Of course, the pure Walecka model is unable to describe nuclear
matter and finite nuclei in a quantitative way. The medium dependence
induced by many-body correlations is not reproduced properly. This
leads to an effective density dependence, which can be taken into
account either by non-linear self-interactions between the mesons or
by a density dependence of the coupling constants. Models with
density dependent coupling constants are tailored more in the spirit
of density functional theory. Indeed, they have turned out to be very
successful~\cite{TW.99,DD-ME1,DD-ME2}. However, they are technically
more complicated, in particular, for RPA calculations. On the other
hand, non-linear models have some justification. In fact, the effects
of vacuum polarization which can contribute up to 30 \% to the
resulting binding energies are not taken into account explicitly in
the framework of the no-sea approximation, but only globally by
adjusting the parameters of the model. A proper renormalization
procedure in nuclear matter \cite{Chin.77} and in finite nuclei
\cite{HS.84,Per.87,ZMR.91} leads to counterterms in the form of
non-linear meson coupling. From this point of view non-linear meson
coupling theories seem at a first glance to be more appropriate for a
phenomenological description of vacuum polarization as compared to
those using density dependent coupling constants. Of course, by
adjustment of the final Lagrangian to the same experimental data,
both methods to introduce a density dependence lead to very similar
results. However the details are different: non-linear meson
couplings lead in the case of nuclear matter to density dependent
meson masses, i.e. to density dependent range parameters of the force
and no rearrangement terms in the Dirac equation, whereas density
dependent coupling constants lead to \ density dependent strenght
parameters of the effective force and additional rearrangement terms.
So far such differences have not been investigated in detail.

The parameter set NL3 \cite{NL3} with non-linear meson couplings
represents one of most successful non-linear RMF forces. It was
proposed ten years ago. In the meantime, new experimental data on
nuclear masses have appeared. Moreover, new and more reliable
information about the neutron skin became available. On the other
hand, it was found that NL3 encounters some difficulties in
describing light Hg and Pb isotopes~\cite{NVRL.02} and certainly,
there is always a need for better predictions of the masses which, of
course, reflect also correct nuclear sizes. For this reason we
decided to improve the parameter set NL3 by performing a new global
fit of ground state properties of spherical nuclei and infinite
nuclear matter. New parameterization obtained in this fit will be
called NL3*.

In order to stay as close as possible to the very successful
parameter set NL3, we kept in this fit the functional form of the
Lagrangian unchanged. In particular the density dependence of the
both parameter sets NL3 and NL3* is determined by the non-linear
meson couplings in the isoscalar channel $\sigma $. We will discuss
the adequacy of this ansatz at the end the this letter.

The starting point of Covariant Density Functional Theory (CDFT) is a
standard Lagrangian density~\cite{GRT.90}
\begin{eqnarray}
\mathcal{L}  &  =&\bar{\psi}\left(
\gamma(i\partial-g_{\omega}\omega-g_{\rho
}\vec{\rho}\vec{\tau}-eA)-m-g_{\sigma}\sigma\right)  \psi\nonumber\\
&&  +\frac{1}{2}(\partial\sigma)^{2}-\frac{1}{2}m_{\sigma}^{2}\sigma^{2}%
-\frac{1}{4}\Omega_{\mu\nu}\Omega^{\mu\nu}+\frac{1}{2}m_{\omega}^{2}\omega
^{2}\label{lagrangian}\\
&&  -\frac{1}{4}{\vec{R}}_{\mu\nu}{\vec{R}}^{\mu\nu}+\frac{1}{2}m_{\rho}%
^{2}\vec{\rho}^{\,2}-\frac{1}{4}F_{\mu\nu}F^{\mu\nu}\nonumber
\end{eqnarray}
which contains nucleons described by the Dirac spinors $\psi$ with
the mass $m$ and several relativistic fields characterized by the
quantum numbers of spin, parity, and isospin. These are effective
fields mediated by mesons, with no direct connection to mesons and
resonances existing in free space. They only carry their quantum
numbers and characterize the properties of the possible relativistic
fields entering the effective Dirac equation, which corresponds to
the Kohn-Sham equation~\cite{KS.65} in the non-relativistic case. It
is only for simplicity that we use the conventional names $\sigma$
($I^{\pi}=0^{+},T=0$), $\omega$ ($I^{\pi}=1^{-},T=0$), and $\rho$
($I^{\pi }=1^{-},T=1$). In addition, we have the electromagnetic
field $A$. In principle there should also be an effective scalar
meson with $I^{\pi}=0^{+},$ and $T=0$, the $\delta$-meson. It plays
an important role in G-matrix calculations ~\cite{SOA.05}. However,
present phenomenological parameterizations of the relativistic energy
density functional~\cite{NL3,DD-ME1,DD-ME2} are very successful
without the inclusion of this meson. On the basis of present
experimental data in finite nuclei it is very difficult to
distinguish between effective scalar and vector mesons in the isospin
$T=1$ channel, and, therefore, all the isospin dependence is carried
by the effective $\rho$-meson.

\begin{table}[t]
\centering%
\tabcolsep=3pt%
\renewcommand{\arraystretch}{1.5}%
\caption{ The total binding energies B.E., charge radii r$_{c}$, and
the differences between the radii of neutron and proton density
distributions r$_{np}$ = (r$_{n}$ $-$ r$_{p}$), used to adjust the
new parameter set NL3*. The calculated values are compared with
experimental data (values in
parentheses).}%
\label{Tab2}
\begin{center}
\bigskip%
\begin{tabular}
[c]{ccll}\hline\hline%
& B.E.(MeV)& r$_{c}$ (fm) & r$_{n}$-r$_{p}$(fm)\\\hline
$^{16}$O  &~-128.112(~-127.619) & 2.735(2.730) & -0.15\\
$^{40}$Ca &~-341.578(~-342.052) & 3.470(3.485) & -0.14\\
$^{48}$Ca &~-413.615(~-415.990) & 3.470(3.484) & ~0.14\\
$^{72}$Ni &~-612.168(~-613.152) & 3.892~~~~~~~ & ~0.26 \\
$^{90}$Zr &~-782.368(~-783.891) & 4.263(4.272) & ~0.04 (0.07)\\
$^{116}$Sn &-986.512(~-988.680) & 4.604(4.626) & ~0.10 (0.12)\\
$^{124}$Sn &-1048.324(-1049.962) & 4.655(4.674) & ~0.22 (0.19)\\
$^{132}$Sn &-1101.550(-1102.850) & 4.636~~~~~~~ & ~0.29\\
$^{204}$Pb &-1608.100(-1607.505) & 5.432(5.486) & ~0.19\\
$^{208}$Pb &-1638.230(-1636.430) & 5.508(5.505) & ~0.23 (0.20)\\
$^{214}$Pb &-1660.119(-1663.290) & 5.566(5.562) & ~0.26\\
$^{210}$Po &-1649.024(-1645.210) & 5.544~~~~~~ & ~0.20\\%
\hline\hline
\end{tabular}
\end{center}
\end{table}

The Lagrangian (\ref{lagrangian}) contains as parameters the meson
masses $m_{\sigma}$, $m_{\omega}$, and $m_{\rho}$ and the coupling
constants $g_{\sigma}$, $g_{\omega}$, and $g_{\rho}$. This model has
first been introduced by Walecka~\cite{Wal.74,SW.86}. It soon became
clear that surface properties of finite nuclei, in particular, the
incompressibility, cannot be described properly by this model.
Therefore, Boguta and Bodmer~\cite{BB.77}
introduced a non-linear meson coupling%
\begin{equation}
U(\sigma)~=~\frac{1}{2}m_{\sigma}^{2}\sigma^{2}+\frac{1}{3}g_{2}\sigma
^{3}+\frac{1}{4}g_{3}\sigma^{4}.
\end{equation}
which brings in an additional density dependence. By fitting
experimental data of some carefully chosen spherical nuclei, very
successful parameter sets have been proposed. For example, the set
NL1~\cite{NL1} is an excellent force for describing ground state and
rotational properties of finite nuclei along the valley of stability.
However, it produces a very large asymmetry parameter $a_{4}$.
Therefore, later, the parameter set NL3~\cite{NL3} was introduced to
improve the asymmetry parameter using the same ansatz without
increasing the number of phenomenological parameters. This set is
able to describe many ground state properties of finite nuclei all
over the periodic table. Recently, the set NL3 has also been applied
very successfully for the investigation of excited states such as
collective rotations~\cite{AKRE.00,AKF.03,AF.05b} and giant
resonances~\cite{VBR.95,VLB.97,VPR.01a,VPR.01b}. In particular, the
use of the effective force NL3 in the $ph$-channel together with the
Brink-Booker part of the Gogny force D1S in the
$pp$-channel~\cite{GEL.96}, within the framework of RHB theory, led
to a very successful tool for the description of many properties of
ground states as well as of excited states with collective character.
The NL3 force was for a long time the only parameterization of the
Lagrangian of the RMF theory that was able to provide simultaneously
a consistent description of both excitation energies of giant
monopole and giant dipole resonances.

In Table~\ref{Tab1}, the experimental values of the energies of the
Giant Monopole (GMR) and Giant Dipole Resonances (GDR) in $^{208}$Pb
are compared with the predictions of various non-linear forces. It is
clearly seen that all forces give reasonable values for the GDR
energy, which are close to experiment. This is not the case, however,
for the giant monopole resonance. There, all forces predict values
which, either overestimate (NL-SH~\cite{NLSH} and NL-RA1~\cite{NLRA})
or underestimate (NL1~\cite{NL1}, NLZ~\cite{NLZ}) the experimental
value. Only NL3 and NL3$^*$ give results which are in excellent
agreement with experiment. This, of course, is connected with the
value of the incompressibility of nuclear matter predicted by each of
these forces.

%%%%%%%%%%%%%%%%%%%%%%%%%%%%%%%%%%%%%%%%%%%%%%%%%
\begin{table}[t]
\caption{{\protect\small Parameters of the effective interaction NL3*
in the RMF theory together with the nuclear matter properties
obtained with this effective force. The values of nuclear matter
properties obtained with NL3 are shown in parentheses.}}%
\label{Tab3}
\begin{center}%
\begin{tabular}
[c]{rlrl}\hline\hline & Parameters of NL3* &  & \\\hline
$M$ & = 939 (MeV) &  & \\
$m_{\sigma}$ & = 502.5742 (MeV) & $g_{\sigma}$ & = 10.0944\\
$m_{\omega}$ & = 782.600 (MeV) & $g_{\omega}$ & = 12.8065\\
$m_{\rho}$ & = 763.000 (MeV) & $g_{\rho}$ & = 4.5748\\
$g_{2}$ & = -10.8093 (fm$^{-1}$) &  & \\
$g_{3}$ & = -30.1486 &  & \\
\  &  &  & \\\hline & Nuclear matter properties &  & \\\hline
$\rho_{0}$ & = 0.150 \qquad(0.148) fm$^{-3}$ &  & \\
$(E/A)_{\infty}$ & = 16.31 \qquad(16.30) MeV &  & \\
$K$ & = 258.28\qquad(271.76) MeV &  & \\
$J$ & = 38.6 ~\qquad(37.4) MeV &  & \\
$m^{*}/m$ & = 0.594 \qquad(0.60) &  & \\\hline\hline
\end{tabular}
\end{center}
\end{table}
%%%%%%%%%%%%%%%%%%%%%%%%%%%%%%%%%%%%%%%%%%%%%%%%%

The starting point for the fit of the new parameter set were the NL3
values, this time, however, the mass $m_{\omega}$ of the
$\omega$-meson was kept fixed to $m_{\omega}=782.6$ MeV. This value
differs slightly from that obtained in the NL3 fit in Ref. \cite{NL3}
and is the same as the one proposed for the free nucleon-nucleon
force by Machleidt in Ref.~\cite{Mac.89}. As shown in
Table~\ref{Tab2} we used twelve spherical nuclei for the present fit: $^{16}%
$O, $^{40}$Ca, $^{48}$Ca, $^{72}$Ni, $^{90}$Zr, $^{116}$Sn,
$^{124}$Sn, $^{204}$Pb, $^{208}$Pb, $^{214}$Pb, and $^{210}$Po.
Compared with the input
for the NL3 fit in Ref. \cite{NL3} we added the nuclei $^{204}$Pb and $^{210}%
$Po and replaced $^{58}$Ni by $^{72}$Ni in order to include more data
on neutron rich isotopes. The experimental binding energies were
taken from Ref.~\cite{AWT.03}, and the charge radii from a recent
compilation~in Ref. \cite{CHARGE}. The most recent experimental
information on the neutron skins
of the nuclei $^{90}$Zr~\cite{YSS.06}, $^{116}$Sn, $^{124}$Sn and $^{208}%
$Pb~\cite{SH.94,KFA.99} was also included. The values in parentheses
correspond to the relative error bars used in the fitting procedure.
The binding energies of finite nuclei and the charge radii are taken
within an uncertainty of 0.1\% and 0.2\%, respectively. The error
used for the neutron skin is 5\%.

For the open shell nuclei pairing correlations are treated in the BCS
approximation with empirical pairing gaps (five-point formula).
Nuclear matter information was also considered in the fit. The
"empirical" input is: $E/A=-16$ MeV (5\%), $\rho_{0}=0.153$ fm$^{-3}$
(10\%), $K_{0}=250$ MeV (10\%), and $J=33$ MeV (10\%). The values in
parentheses correspond to the error bars used in the fitting
procedure. The resulting values of the Lagrangian parameterization
NL3* are given in Table~\ref{Tab3}, together with the predictions of
the force for the nuclear matter properties at the saturation point.
The values in parentheses are the predictions obtained with the
effective interaction NL3. The resulting energy at saturation and
asymmetry energy are rather close to the former values, only
incompressibility is about 12 MeV smaller. Nonetheless both values
are still sowewhat to high as compared with other paremeter sets with
a density dependence in the iso-vector channel \cite{Pie.02,VNR.03}.

%%%%%%%%%%%%%%%%%%%%%%%%%%%%%%%%%%%%%%%%%%%%%%%%%%%%%%%%%%%%%%%%%%%%%%%%%%%%%%%%
\begin{figure}[ptb]
\centering
\includegraphics[width=8cm]{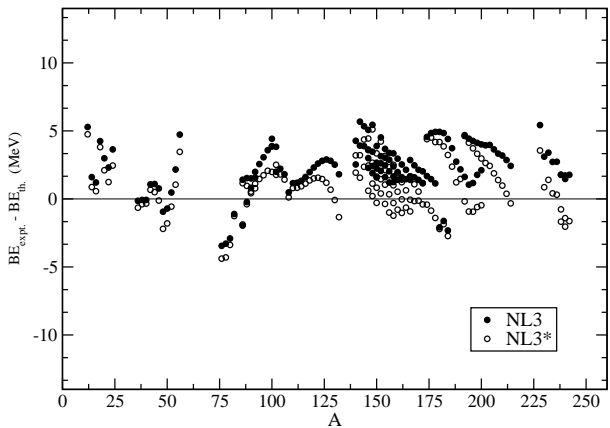} \newline%
\caption{{\protect\small Absolute deviations of the binding energies
calculated with the parameter sets NL3 (filled circles) and NL3*
(open circles) from the experimental values of Ref.~\cite{AWT.03}. }}%
\label{figA}%
\end{figure}
%%%%%%%%%%%%%%%%%%%%%%%%%%%%%%%%%%%%%%%%%%%%%%%%%%%%%%%%%%%%%%%%%%%%%%%%%%%%%%%%

In Fig.~\ref{figA} the binding energies of more than 180 even-even
nuclei are compared with experiment and the predictions of the NL3
forces. All calculations have been performed within the RHB model
with the Gogny force D1S ~\cite{BGG.84} in the pairing channel.
\begin{equation}
V^{pp}(1,2)~=~\sum_{i=1,2}e^{-\frac{r^2}{\mu_{i}^{2}}%
}\,(W_{i}~+~B_{i}P^{\sigma}-H_{i}P^{\tau}-M_{i}P^{\sigma}P^{\tau}).
\end{equation}
The parameters $\mu_{i}$, $W_{i}$, $B_{i}$, $H_{i}$, and $M_{i}$
$(i=1,2)$ of this force are given in Ref.~\cite{BGG.91}. It should be
emphasized that, since a finite-range pairing interaction is used,
the results do not depend on unphysical parameters like, the momentum
cut-off in the pairing channel.

The results of NL3* are shown as open circles while those obtained
with NL3 are marked by filled circles. For light nuclei, both forces
give similar predictions, however, as the mass number increases NL3*
results are clearly closer to the zero MeV line.

\begin{figure}[t]
\centering
\includegraphics[width=8cm]{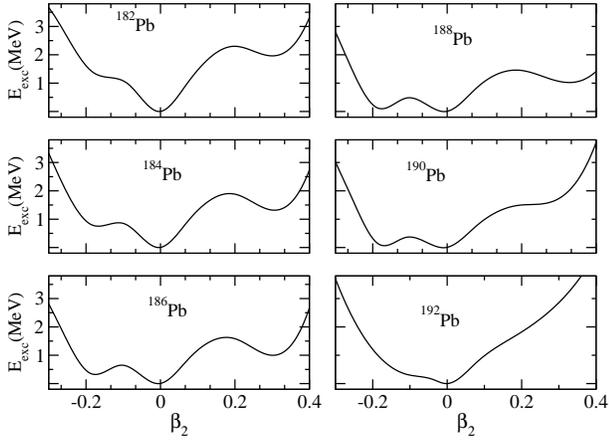} \newline%
\caption{{\protect\small Excitations energies of even-Pb isotopes as
a
function of the deformation parameter $\beta_{2}$.}}%
\label{figB}%
\end{figure}

As discussed above the parameter set NL3 had difficulties to
reproduce the proper ground state deformations in light Hg and Pb
nuclei~\cite{NVRL.02}. This is no longer the case with the parameter
set NL3*. To investigate this, we have carried out constrained
axially deformed RHB calculations of several even-A Pb isotopes with
masses between $182\leq A\leq192$ in an external quadrupole field and
we display in Fig.~\ref{figB} the corresponding energy surfaces as a
function of the quadrupole deformation. It is seen that in all cases
the Pb isotopes turn out to be spherical. This is also the case for
all other Pb isotopes, which are not shown in the figure. It is also
seen that the Pb isotopes manifest the interesting effect of shape
coexistence. The energies which correspond to the oblate and prolate
shape solutions are very close to the spherical ones but definitely
lay higher in energy. This indicates a clear improvement as compared
to the parameter set NL3, where some light Pb isotopes showed a
deformed shape~\cite{NVRL.02}. A more quantitative analysis goes
beyond the mean field limit and requires, for instance,
GCM-calculations \cite{NVR.06a,NVR.06b}. It is essential, however,
that the mean field solution, which is the starting point for such
investigations gives the correct behavior.

\begin{figure}[ptb]
\centering
\includegraphics[width=8cm]{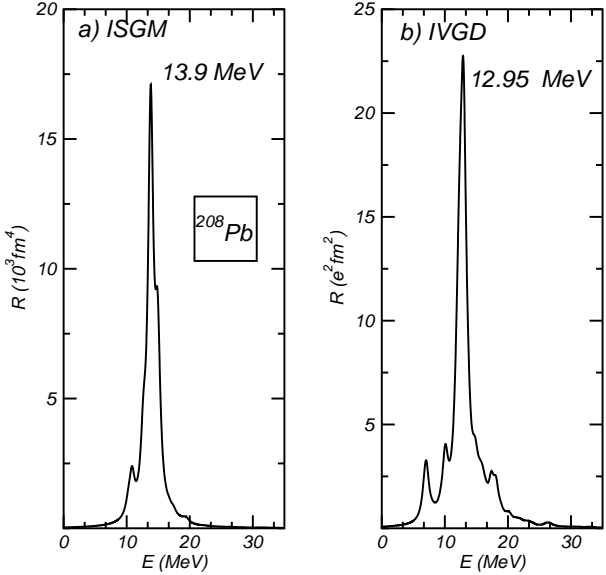}\caption{{\protect\small The
isoscalar monopole (a), and the isovector dipole (b) strength
distributions in $^{208}$Pb calculated with the effective interaction
NL3*. The experimental excitation energies are $14.1\pm0.3$
MeV~\cite{YCL.99} for the monopole resonance, and $13.3\pm0.1$
MeV~\cite{RBK.93} for the dipole resonance,
respectively.}}%
\label{figC}%
\end{figure}

In the following we investigate dynamical processes such as
collective vibrations with the same parameter set NL3*. For that
purpose we study the time-dependent RMF or RHB equations in the small
amplitude limit \cite{RMG.01}, i.e. we solve the relativistic RPA or
QRPA equations. In Fig.~\ref{figC} we display results for the
monopole and iso\-vector dipole response for the nucleus $^{208}$Pb.
For the multipole operator $\hat {Q}_{\lambda\mu}$ the response
function $R(E)$ is defined
\begin{equation}
R(E)=\sum_{i}~B(J_{i}\rightarrow0_{f})~{\frac{{\Gamma/{2\pi}}}{{(E-E_{i}%
)^{2}+\Gamma^{2}/4}}}, \label{Lorentz}%
\end{equation}
where $\Gamma$ is the width of the Lorentzian distribution, and
\begin{equation}
B(J_{i}\rightarrow0_{f})=\frac{1}{2J_{i}+1}\,|\langle
J_{i}||\hat{Q}_{\lambda }||0_{f}\rangle|^{2}.
\end{equation}
In the examples considered here the continuous strength distributions
are obtained by folding the discrete spectrum of R(Q)RPA states with
the Lorentzian (see Eq. (\ref{Lorentz})) with constant width
$\Gamma=0.5$ MeV~\cite{VPR.01a}. The calculated peak energies of the
ISGMR resonance at 13.9 MeV and of the IVGDR resonance at 12.95 MeV
should be compared with the experimental excitation energies:
$E=14.1\pm0.3$ MeV~\cite{YCL.99} for the monopole resonance, and
$E=13.3\pm0.1$ MeV~\cite{RBK.93} for the dipole resonance,
respectively. Clearly, the agreement with experiment is very good.
The NL3 force predicts the same value for the IVGDR and a slightly
larger value for the ISGMR. This is to be expected as the predicted
value for the nuclear incompressibility with NL3 is larger than the
one obtained with NL3*. However, both values are within the
experimental error bars and one could say that RMF theory with
non-linear meson couplings leads to values of the incompressibility
around 260 $\pm10$ MeV.
%%%%%%%%%%%%%%%%%%%%%%%%%%%%%%%%%%%%%%%%%%%%%%%%%%%%%%%%%%%%%%%%%%%%%%%%%%%%%%%%
%%%%%%%%%%%%%%%%%%%%%%%%%%%%%%%%%%%%%%%%%%%%%%%%%%%%%%%%%%%%%%%%%%%%%%%%%%%%%%%%

%%%%%%%%%%%%%%%%%%%%%%%%%%%%%%%%%%%%%%%%%%%%%%%%%%%%%%%%%%%%%%%%%%%%%%%%%%%%%%%%
%%%%%%%%%%%%%%%%%%%%%%%%%%%%%%%%%%%%%%%%%%%%%%%%%%%%%%%%%%%%%%%%%%%%%%%%%%%%%%%%

\begin{figure}[ptb]
\centering
\includegraphics[width=7cm]{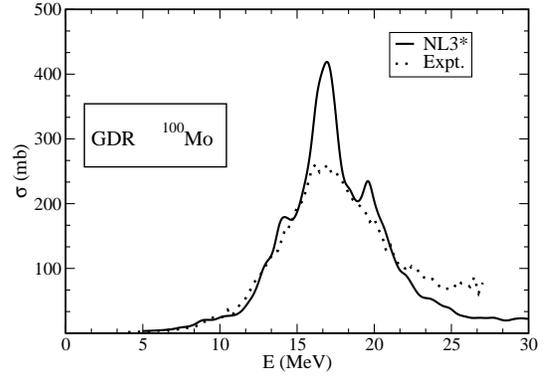}\caption{{\protect\small The total
isovector electric dipole cross section (in mb) of the deformed
nucleus $^{100}$Mo as a function of the excitation energy, calculated
with the
effective force NL3*, is compared with experiment {\cite{BBC.74,RGE.06}.}}}%
\label{figD}%
\end{figure}

Recently a new computer code has been developed for the solution of
the relativistic QRPA equations in axially deformed nuclei
\cite{Pen.07,PR.08}. We used this code for the study of giant
resonances in deformed nuclei. Here, we present as an example
calculations in the prolate deformed nucleus $^{100}$Mo. In
Fig.~\ref{figD} we show the total isovector dipole cross section as
function of the GDR energy. The parameter set NL3* is very effective
in reproducing these experimental data. The full line corresponds to
the fully self-consistent deformed relativistic QRPA calculations
while the dotted line are the experimental data~\cite{BBC.74,RGE.06}.
The estimated centroid energy for the GDR differs from the
experimental value by less than 0.2 MeV. It is noted that NL3 also
predicts excellent results, however, our analysis shows that the
results with the newly developed NL3* are slightly better. This can
be traced back to the improvement in the density dependence of the
non-linear sigma channel of the new force.

%%%%%%%%%%%%%%%%%%%%%%%%%%%%%%%%%%%%%%%%%%%%%%%%%%%%%%%%%%%%%%%%%%%%%%%%%%%%%%%%
%%%%%%%%%%%%%%%%%%%%%%%%%%%%%%%%%%%%%%%%%%%%%%%%%%%%%%%%%%%%%%%%%%%%%%%%%%%%%%%%

%%%%%%%%%%%%%%%%%%%%%%%%%%%%%%%%%%%%%%%%%%%%%%%%%%%%%%%%%%%%%%%%%%%%%%%%%%%%%%%%
\begin{figure*}[t]
\centering
\includegraphics[width=15cm]{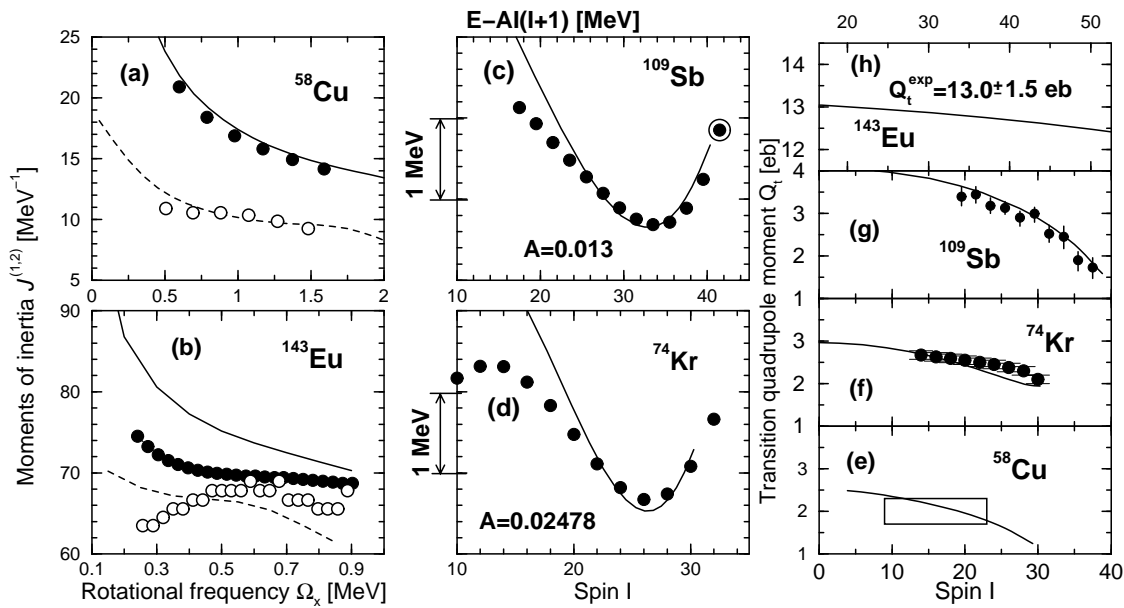}\caption{{\protect\small Panels
(a-b):. Kinematic (solid circles, solid lines) and dynamic (open
circles, dashed lines) moments of inertia in the highly-deformed band
in $^{58}$Cu and the yrast superdeformed band of $^{143}$Eu.
Experimental data are shown by circles, while the results of
calculations by lines. Panels (c-d): Excitation energies of the
non-terminating band in $^{74}$Kr and the smoothly terminating band
in $^{109}$Sb relative to a rigid rotor reference $E_{RLD}(I)=A
I(I+1)$, where $A$ is a moment of inertia parameter. Panels (e-h):
Experimental and calculated transition quadrupole moments $Q_{t}$.
Only the data for $^{143}$Eu include the errors due to the
uncertainties in the nuclear and electronic stopping powers. Other
data exclude these uncertainties which are on the level of 10-15\% of
absolute value of $Q_{t}$. The box in panel (e) indicates the upper
and lower limits of $Q_{t}$ and the spin range where they have been
measured. The panel (h) itself acts as such a box in the case of the $^{143}%
$Eu band. Experimental data are taken from Refs.~\cite{RBD.98}
($^{58}$Cu), \cite{VSS.05} ($^{74}$Kr), \cite{SLF.96,WCC.98}
($^{109}$Sb), and
\cite{FAH.95} ($^{143}$Eu).}}%
\label{figE}%
\end{figure*}
%%%%%%%%%%%%%%%%%%%%%%%%%%%%%%%%%%%%%%%%%%%%%%%%%%%%%%%%%%%%%%%%%%%%%%%%%%%%%%%%

The response of the nuclei to an external force provides an important
test of our understanding of their microscopic structure. The
Coriolis force is one of such forces. It is of particular interest
because it breaks time reversal symmetry in the intrinsic frame and
leads therefore to currents and to nuclear magnetic fields having its
origin in the spatial contributions of the vector mesons
\cite{KR.89}. The impact of this field can be studied in rotating
nuclei. Such studies were previously performed with various
non-linear RMF parameterizations in the different parts of the
nuclear chart, see Ref.~\cite{VALR.05} and references therein. In
order to test the parameter set NL3* \ we selected several
representative examples. This choice was guided by (i) the necessity
to test the accuracy of the description of different types of
rotational bands in different mass regions, and (ii) the restriction
to consider only bands which are very weakly affected by the pairing
correlations. The later constraint is related to the fact that the
pairing modifies considerably the rotational properties and some
inadequacies in the pairing channel may prevent meaningful
conclusions about the reliability of the parameterization. For a
discussion of this point see Ref.~\cite{AKF.03}. Thus, a
highly-deformed band in the nucleus $^{58}$Cu~\cite{RBD.98}, a
non-terminating ground state band in $^{74}$Kr~\cite{VSS.05}, \ a
smoothly-terminating band in $^{109}$Sb~\cite{SLF.96}, and an yrast
superdeformed (SD) band in $^{143}$Eu~\cite{FAH.95} were selected for
this comparison. The pairing correlations are weak in these bands,
and thus all calculations were performed within the framework of
cranked relativistic mean field (CRMF) theory~\cite{AKR.96} which
neglects pairing.

The results of these calculations for rotational and deformation
properties of the bands mentioned before are compared with experiment
in Fig.~\ref{figE}. The kinematic ($J^{(1)}$) and dynamic ($J^{(2)}$)
moments of inertia (Fig. 5a) of the highly-deformed band in $^{58}$Cu
as well as its transition quadrupole moment $Q_{t}$ (Fig. 5e) are
well described by the CRMF calculations. The fact that the $J^{(1)}$
moment is considerably larger than the $J^{(2)}$ moment and that both
moments smoothly decrease with rotational frequency are clear
fingerprints that the pairing is not important at high
spin~\cite{VALR.05}. While this condition is clearly satisfied in
$^{58}$Cu (see Fig. 5a), these moments are closer to each other in
the case of the SD band in $^{143}$Eu (Fig. 5b) suggesting that the
pairing is relatively more important in the later band. As a result,
we observe somewhat larger discrepancies between experiment and
theory in the case of $^{143}$Eu (Fig. 5b). If the calculations would
include weak pairing, this would lead to the decrease (increase) of
calculated $J^{(1)}$ ($J^{(2)}$) moments of inertia bringing them
closer to experimental data. It is also necessary to mention that the
SD band in $^{143}$Eu is not linked to the low-spin spectra.

Based on the effective alignment approach with respect of the linked
yrast SD band in $^{152} $Dy~\cite{ALR.98,VALR.05} the spin of the
initial state of this band has been suggested to be

$I_{0}=16.5\hbar$: this value has been used in the plotting of
experimental kinematic moment of inertia in Fig. 5b. However, one
cannot completely exclude also the value $I_{0}=18.5\hbar$ (see
discussion in Ref.~\cite{AKR.96}); this would increase experimental
$J^{(1)}$ values bringing them closer to calculated ones. The
rotational properties of the bands which come close to their maximum
spin $I_{max}$ (as defined from the distribution of particles/holes
over low- and high-$j$ orbitals) are usually analyzed employing
excitation energies plotted versus rigid rotor reference. This type
of analysis is used for the bands in $^{109}$Sb and $^{74}$Kr, see
Figs. 5c and 5d. One can see that at high spin above $I\sim20\hbar$,
where the pairing is expected to be weak, the CRMF results describe
the excitation energies very well. The band in $^{109}$Sb is a
classical example of a smoothly terminating band; it starts at low
spin at near-prolate shape, then with increasing spin its shape
becomes more and more triaxial and finally at $I_{max}=41.5\hslash$
it terminates in a non-collective oblate state~\cite{VALR.05}. This
process of smooth band termination is associated with the gradual
loss of collectivity which is nicely seen in the drop of the
transition quadrupole moment $Q_{t}$ with increasing spin (see Fig.
5g). This feature is nicely reproduced in the calculations. On the
contrary, the ground state band in $^{74}$Kr reveals a new phenomenon
of non-termination of rotational bands (see Ref.~\cite{VSS.05} for
details): it remains collective at $I_{max}=32\hslash$. Contrary to
the case of the band in $^{109}$Sb, its transition quadrupole moment
is more constant as a function of spin (see Fig. 5f).

One can see that the CRMF calculations with the parameter set NL3*
provide a very successful description of different aspects of
rotating nuclei. In general, the results of the calculations are very
close to the ones obtained earlier with the NL3 parameterization of
the RMF Lagrangian.

As mentioned above, the density dependence of the parameter sets NL3
and NL3* is determined by the non-linear meson couplings in the
isoscalar channel $\sigma$. The vector fields $\omega$ and $\rho$, in
particular, the isovector channel described by the $\rho$-meson are
still coupled linearly and this leads to relatively large neutron
radii and a very stiff equation of state for neutron
matter~\cite{DD-ME1} as well as very stiff symmetry energies as a
function of the density~\cite{VNR.03}. Models with non-linear
couplings in the $\rho$-channel as for instance the parameter set
FSUGold introduced by Piekarewicz in Ref. \cite{TP.05} can improve
this, however at the cost of additional parameters.  Density
dependent covariant functionals eliminate the non-linear coupling
terms and use, instead of that, at the cost of additional parameters,
density dependent coupling constants $g_{\sigma}(\rho)$,
$g_{\omega}(\rho)$, and $g_{\rho}(\rho)$ which depend on the baryon
density $\rho$. This concept is evidently closer to the idea of
density functional theory. On the other hand, one has to keep in mind
that a proper treatment of the effects of vacuum polarization by a
renormalization procedure would require counterterms which are
nonlinear in the meson fields with the same polynomial
structure~\cite{Chin.77,HS.84,Per.87,ZMR.91}. In addition, because of
the relatively small surface diffuseness in nuclear systems and the
narrow range, in which the density is changing from its value at
saturation to zero, calculations of matrix elements with density
dependent coupling constants and of the corresponding rearrangement
terms require a considerably larger numerical effort, in particular,
in the case of axially symmetric and triaxial applications.
Considering all these points the non-linear version of RMF theory is
still a very useful concept because it combines simplicity with
excellent predictions for many nuclear properties.

In conclusion, in the present work we have reconsidered the well
known parameter set NL3 after ten years and by a new fit with modern
experimental data we introduced an improved parameterization NL3* for
the RMF model, which contains only six phenomenological parameters.
It is able to improve the description of nuclear masses and to cure
some small problems observed previously with the NL3 force. At the
same time, it provides excellent results for collective properties of
vibrational and rotational character.

\section*{Acknowledgements}

This work has been supported in part by the Gesellschaft f\"{u}r
Schwerionenforschung (GSI), by the BMBF (Germany) under the project
06 MT246, by the DFG cluster of excellence \textquotedblleft Origin
and Structure of the Universe\textquotedblright\
(www.universe-cluster.de), by the Greek Ministry of National
Education and Religious Affairs, in the framework of the Pythagoras
II, EPEAEK II and E.U. project\ 80861, by the Hellenic State
Scholarship foundation (IKY) and by the Department of Energy under
award number DE-FG02-07ER41459.

%\section*{References}

%\leftline{\bf REFERENCES}
%\bibliographystyle{c:/b/a00/prsty}
%\bibliography{c:/b/a00/refring}

\end{document}